\documentclass[conference]{IEEEtran}

\ifCLASSINFOpdf
\usepackage[pdftex]{graphicx}
\usepackage{graphicx}
\else
\fi
\usepackage{graphicx}
\usepackage{wrapfig}
\usepackage{balance}
\usepackage{cite}
\usepackage{float}
\usepackage{amsmath}
\usepackage[utf8]{inputenc}
\usepackage[final]{pdfpages}
\usepackage{pdfpages}
\usepackage{lipsum}
\usepackage{textcase}
\usepackage{url}
\usepackage{amsmath,esint}
\usepackage{epstopdf}
\usepackage{array}
\usepackage{color}
\usepackage{subcaption}
\usepackage{multicol}

\usepackage{bm}

\DeclareRobustCommand{\uvec}[1]{{%
  \ifcsname uvec#1\endcsname
     \csname uvec#1\endcsname
   \else
    \bm{\hat{\mathbf{#1}}}%
   \fi
}}

	\newcommand\blfootnote[1]{%
		\begingroup
		\renewcommand\thefootnote{}\footnote{#1}%
		\addtocounter{footnote}{-1}%
		\endgroup
	}
\newcolumntype{P}[1]{>{\centering\arraybackslash}p{#1}}
\newcolumntype{M}[1]{>{\centering\arraybackslash}m{#1}}

\pdfoutput=1

\begin{document}
	
\title{Impact of 3D UWB Antenna Radiation Pattern on Air-to-Ground Drone Connectivity 
}

\author{
\IEEEauthorblockN{Jianlin Chen, Devin Raye, Wahab Khawaja, Priyanka Sinha, and Ismail Guvenc 
}
\IEEEauthorblockA{\IEEEauthorrefmark{1}Department of Electrical and Computer Engineering, North Carolina State University, Raleigh, NC}
Email: \{jchen42, dnraye, wkhawaj, psinha2, iguvenc\}@ncsu.edu, 	
}

\maketitle

\blfootnote{This work has been supported in part by NASA under the Federal Award 
NNX17AJ94A, and by an REU supplement for NSF CNS-1453678.}

\begin{abstract}
Three dimensional (3D) radiation pattern of an antenna mounted at a drone can significantly influence the air-to-ground (A2G) link quality. Even when a drone transmitter is very close to a ground receiver, if the antenna orientations are not aligned properly, a  significant degradation can be observed in the received signal power at the receiver. To characterize such effects for a doughnut-shaped antenna radiation pattern, using an ultra-wideband (UWB) transmitter at the drone and a UWB receiver at the ground,  we carry out A2G channel measurements to capture the link quality at the ground receiver for various link distances, drone heights, and antenna orientations. We develop a simple analytical model to approximate the influence of 3D antenna patterns on the received signal strength (RSS), which show reasonable agreement with measurements despite the simplicity of the model and the complicated 3D radiation from the UWB antennas. We also explore how the signal strength can be improved when multiple antennas with different orientations are utilized at transmitter/receiver.       


\end{abstract}

\IEEEpeerreviewmaketitle


    

\section{Introduction}

Drones are expected to be used widely in the close future for various applications including delivery, surveillance, search and rescue, and mobile hot spots. Certain use cases of drones are expected to also require broadband connectivity with ground receivers. Due to wide range of three dimensional (3D)  geometries that the drones will communicate with ground nodes and with each other, understanding the influence of the 3D antenna radiation patterns at drones and ground receivers on link quality carries major importance~\cite{LTEUAV}. 

There has been several recent works in the literature which study the A2G propagation channel characteristics~\cite{khawaja2018survey,matolak2015unmanned,amorim2017radio}, including the efforts by 3GPP on aerial propagation channels for LTE systems~\cite{RP170779,LTEUAV}. However, such studies are typically limited to relatively narrow-band propagation channels.  
A2G channel propagation is studied  in~\cite{khawaja2016uwb} for UWB systems and in~\cite{khawaja2017uav} for mmWave systems; however, impact of antenna radiation patterns has not been taken into account. 
Apart from few studies on the non-monotonic air-to-ground (A2G) link quality behavior due to antenna radiation~\cite{LTEUAV,6566747,6162389}, there are limited experimental/analytical studies that provide insights on the impact of 3D antenna gains on A2G propagation channel.      

\begin{figure}[!t]
	\centering
    \includegraphics[width=6.5cm]{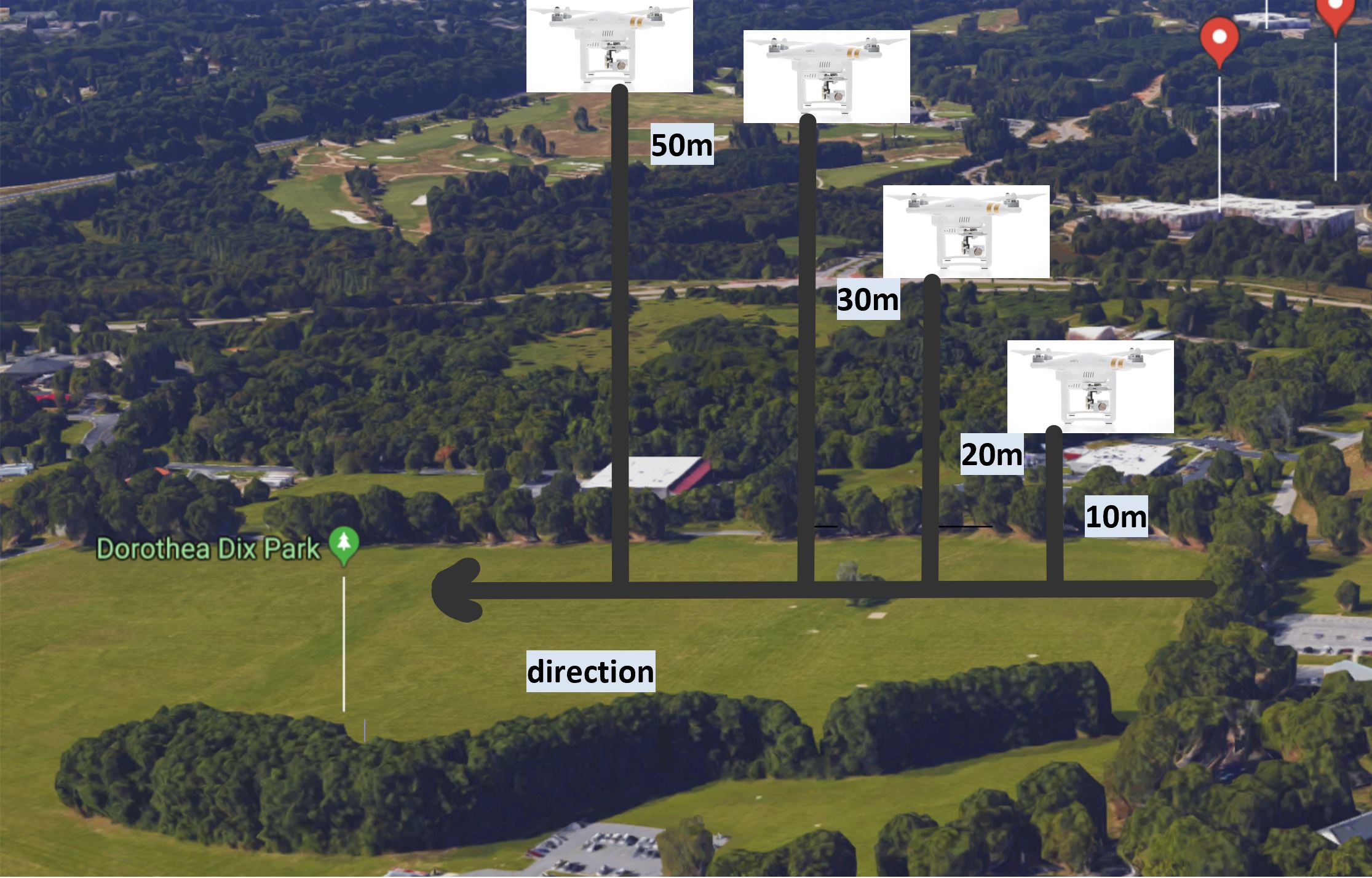}
    \caption{Environment where A2G measurements were collected.}\label{Fig:DroneEnvir}
    \vspace{-5mm}
\end{figure}

In this paper, using an ultra-wideband (UWB) transmitter at the drone and a UWB receiver at the ground (see Fig.~\ref{Fig:DroneEnvir}), both equipped with UWB antennas having doughnut-shaped radiation patterns, we provide results from A2G channel measurements to capture the link quality at the ground receiver for various link distances, drone heights, and antenna orientations. We develop a simple analytical model to approximate the influence of 3D antenna patterns on the received signal strength (RSS), which shows reasonable agreement despite the simplicity of the model and the complicated 3D radiation from the UWB antennas. 
Finally, we explore how much the signal strength can be improved when we utilize multiple antennas with different orientations for an A2G link. 


\section{UWB Air-to-Ground Measurement Setup}

\begin{figure}[h]
	 \vspace{-3mm}
     \begin{multicols}{3}
	\centering
    \begin{subfigure}[b]{0.1\textwidth}
    	\includegraphics[width=\textwidth]{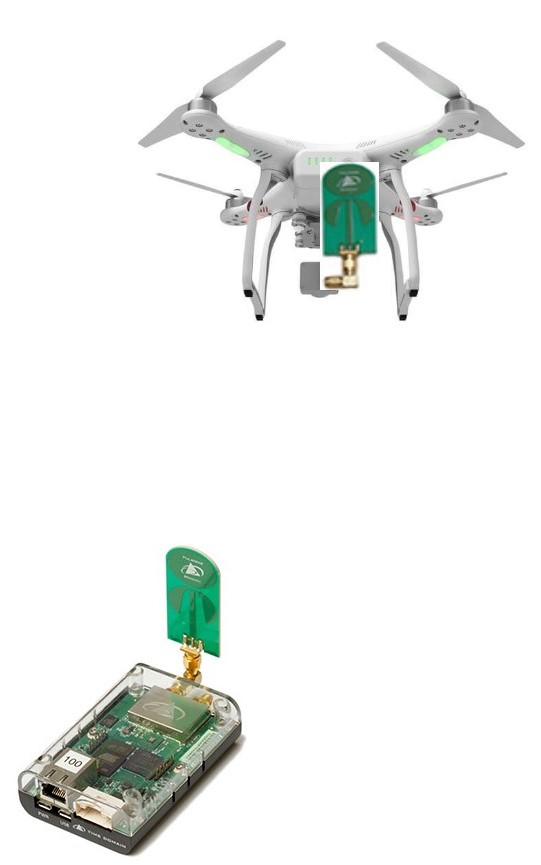}(a)
    \end{subfigure}
    
    \begin{subfigure}{0.1\textwidth}
    \includegraphics[width=\textwidth]{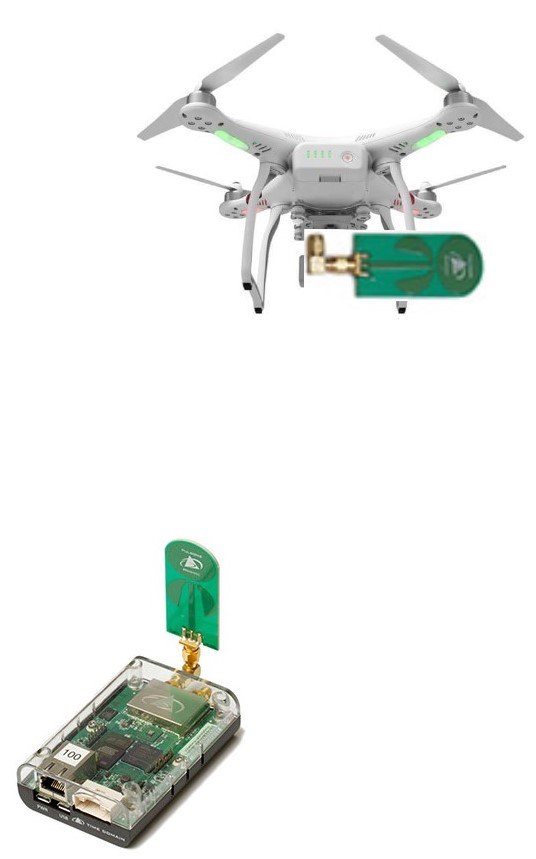}(b)
    \end{subfigure}
    
    \begin{subfigure}{0.1\textwidth}
    \includegraphics[width=\textwidth]{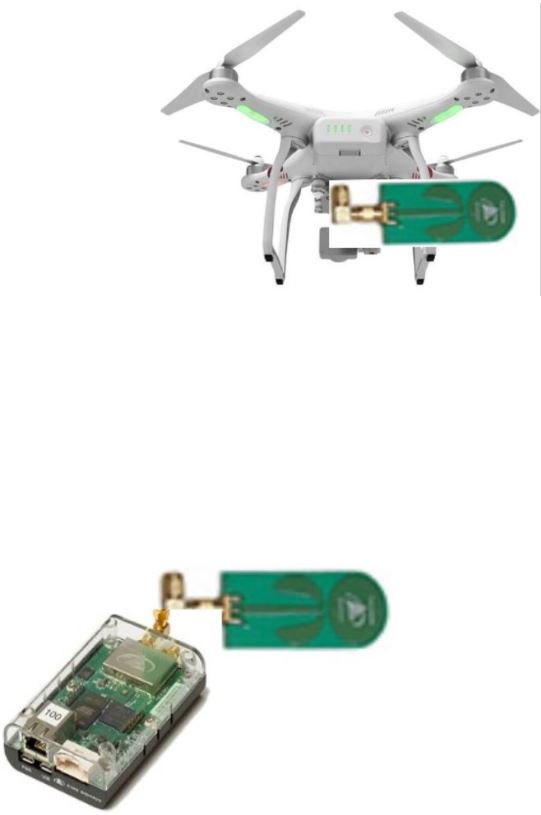}(c)
    \end{subfigure}
    \end{multicols}\vspace{-3mm}
    \caption{Experiments with (a) vertical-vertical (VV), (b) vertical-horizontal (VH), and (c) horizontal-horizontal (HH) configurations for the ground and aerial antennas.}\label{Fig:Orientations}\vspace{-2mm}
\end{figure}

First, we describe our measurement setup for UWB A2G channel propagation experiments. 
We utilize PulsON 440 (P440) kits from Time Domain, Inc., which has an instantaneous bandwidth of 2.2~GHz between 3.1~GHz to 5.3~GHz. 
Using a P440 kit mounted on a tri-pod (roughly 1.27 m off the ground) and a kit attached to a DJI Phantom~3 drone, we were able to determine the RSS and the signal-to-noise ratio (SNR) of the drone flying in a straight line away from the ground unit at a height of 10~m, 20~m, 30~m, and 50~m. For experiments, we considered three different horizontal and vertical orientations of the transmit/receive antennas as illustrated in Fig.~\ref{Fig:Orientations}.  

\begin{figure}[h!]
\centering
\begin{subfigure}{0.4\textwidth}
	\includegraphics[width = \columnwidth]
    {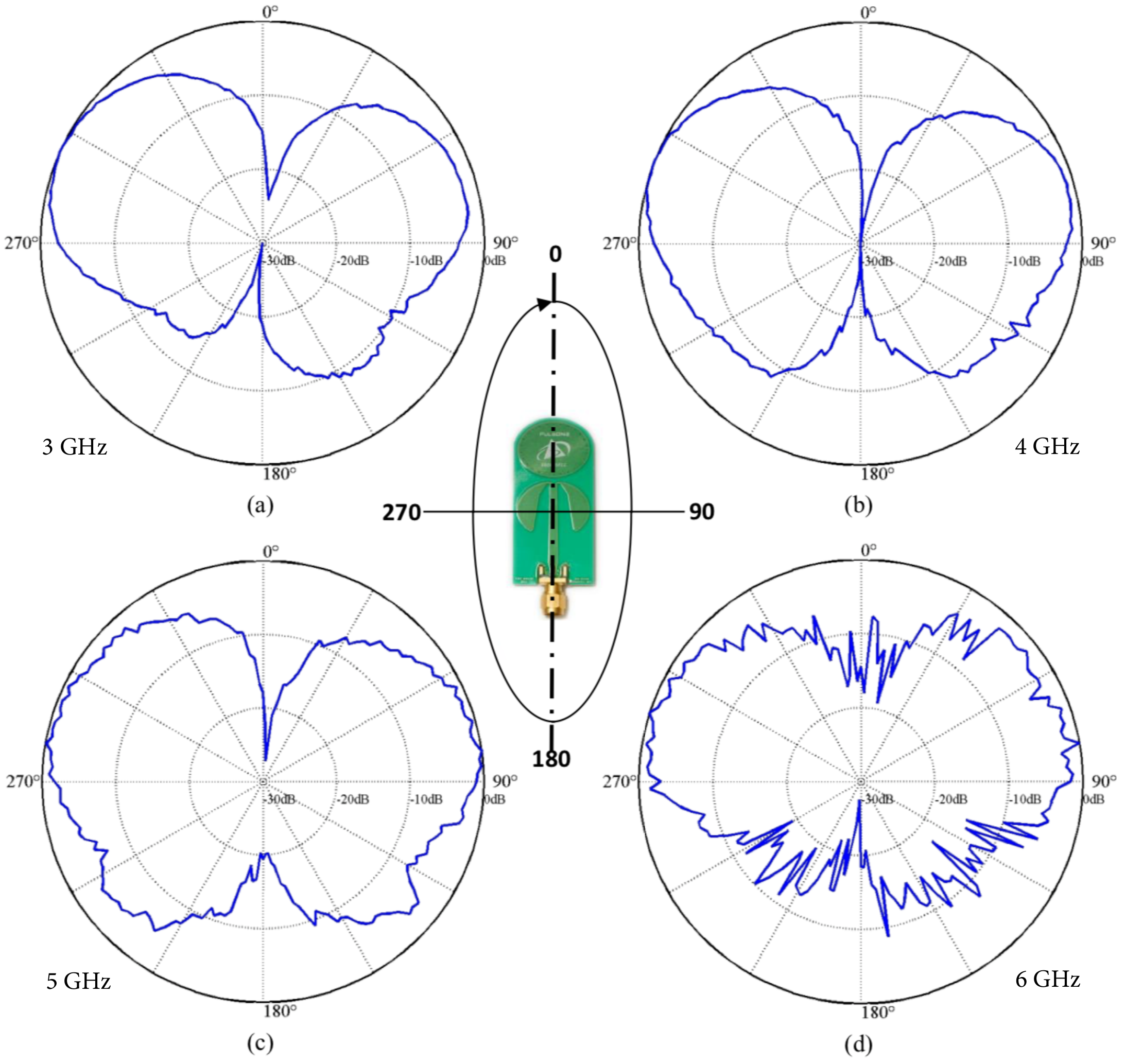}
    \caption{Vertical radiation pattern.}  
\end{subfigure}
    \begin{subfigure}{0.4\textwidth}
    \includegraphics[width = \columnwidth]
    {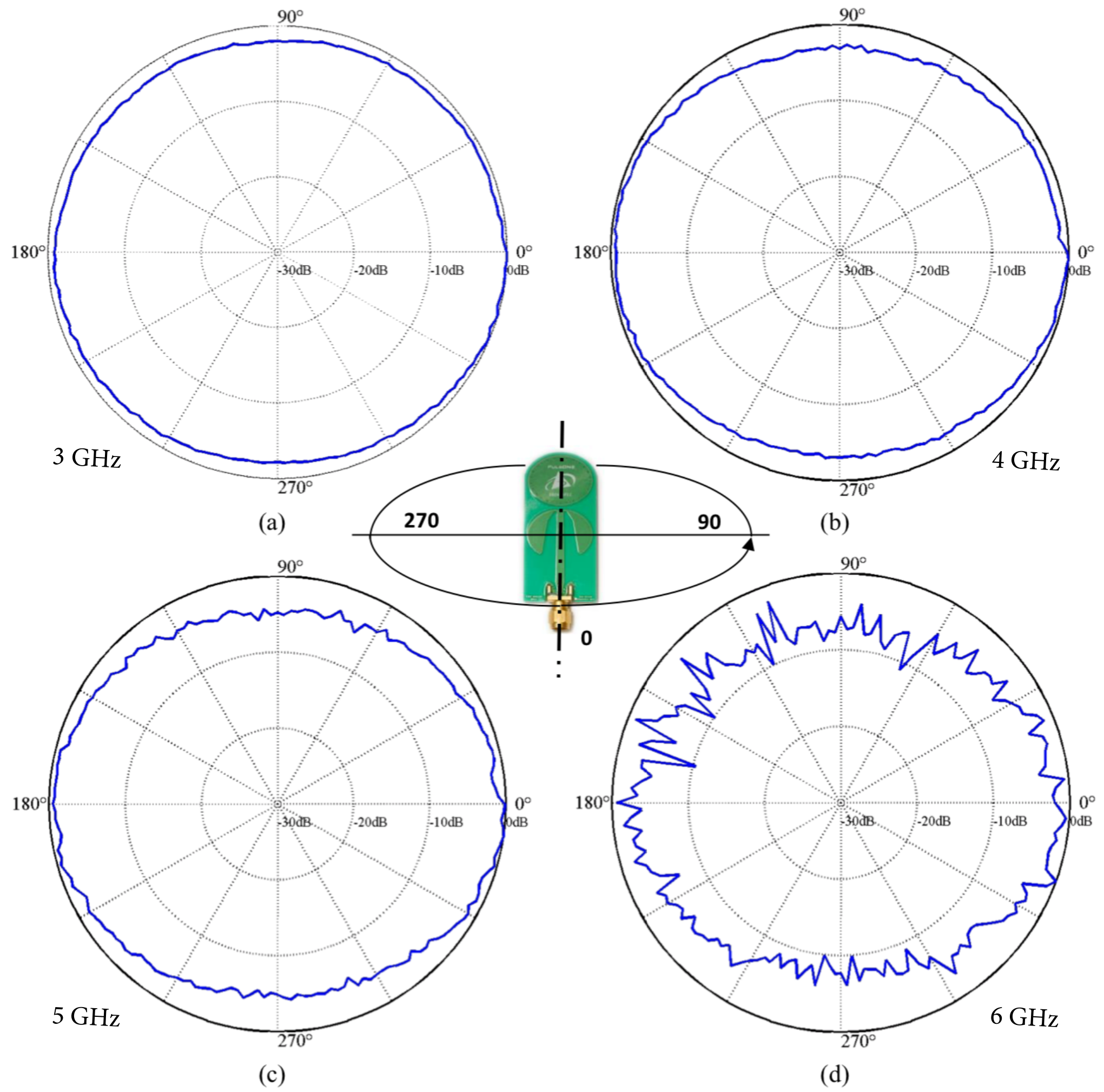}
    \caption{Horizontal radiation pattern.}
     \end{subfigure}
    \caption{Antenna radiation patterns for UWB BroadSpec antenna at 3~GHz, 4~GHz, 5~GHz, and 6~GHz: (a) vertical and (b) horizontal planes~\cite{BroadSpecAntenna}.}\label{Fig:RadiationPattern}
\vspace{-5mm}
\end{figure}


We utilize BroadSpec UWB antennas from Time Domain, Inc., where the horizontal and vertical radiation patterns at different frequency bands of 3~GHz, 4~GHz, 5~GHz, and 6~GHz, are illustrated in Fig.~\ref{Fig:RadiationPattern} from the manufacturer's specifications~\cite{BroadSpecAntenna}. As Fig.~\ref{Fig:RadiationPattern}(a) suggests, the vertical beam pattern varies significantly as a function of elevation angle between a transmitter and the receiver, which (for a fixed drone height) will be changing as a function of the distance between the drone and the ground receiver. While the azimuth radiation pattern is relatively uniform, up to few decibels of power difference may exist depending on the azimuth angle between a transmitter and a receiver. In the rest of this paper, we will assume for simplicity that the antenna gain is independent of the azimuth angle between the transmitter/receiver pair, and we will attempt to model the impact of elevation angle on the antenna gain, and hence the RSS at the ground receiver.   




\section{Modeling the Effect of 3D Antenna Gain}

\begin{figure}[!t]
    \includegraphics[width=0.95\columnwidth]{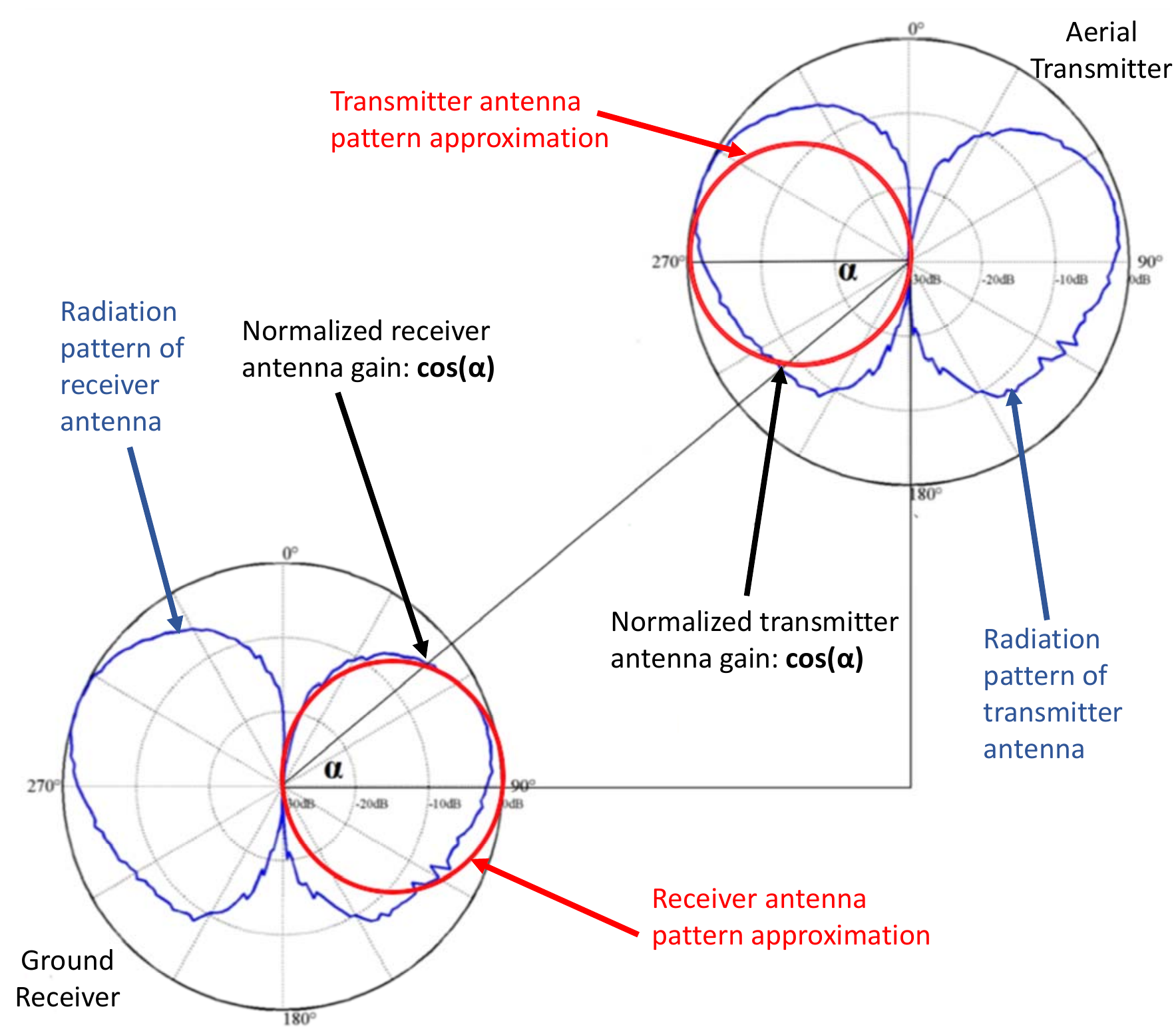}
    \caption{Analytical model for the antenna gains considering vertical-vertical antenna configuration in Fig.~\ref{Fig:Orientations}.}\label{Fig:GeometricModel}\vspace{-3mm}
\end{figure}

Considering a link model as in Fig.~\ref{Fig:GeometricModel}, we can represent the RSS at a ground receiver from an aerial UWB transmitter as: 
\begin{equation}
P_{\rm RX}(\alpha) = P_{\rm TX}G_{\rm TX}(\alpha)G_{\rm RX}(\alpha)\left(\frac{\lambda}{4\pi d}\right)^{\gamma}~,\label{Eq_1}
\end{equation}
where $P_{\rm TX}$ is the transmitted signal power, $G_{\rm TX}(\alpha)\leq 1$ is the transmitter antenna gain, $G_{\rm RX}(\alpha)\leq 1$ is the receiver antenna gain, $\alpha$ and $d$ are the angle and the distance between the transmitter and the receiver, $\lambda$ is the  wavelength, and $\gamma$ is the path loss exponent, which is assumed as two in the rest of this paper due to line-of-sight experimental scenarios.

\begin{wrapfigure}{r}{2in}
\centering \vspace{-3mm}
    \includegraphics[width=2in]{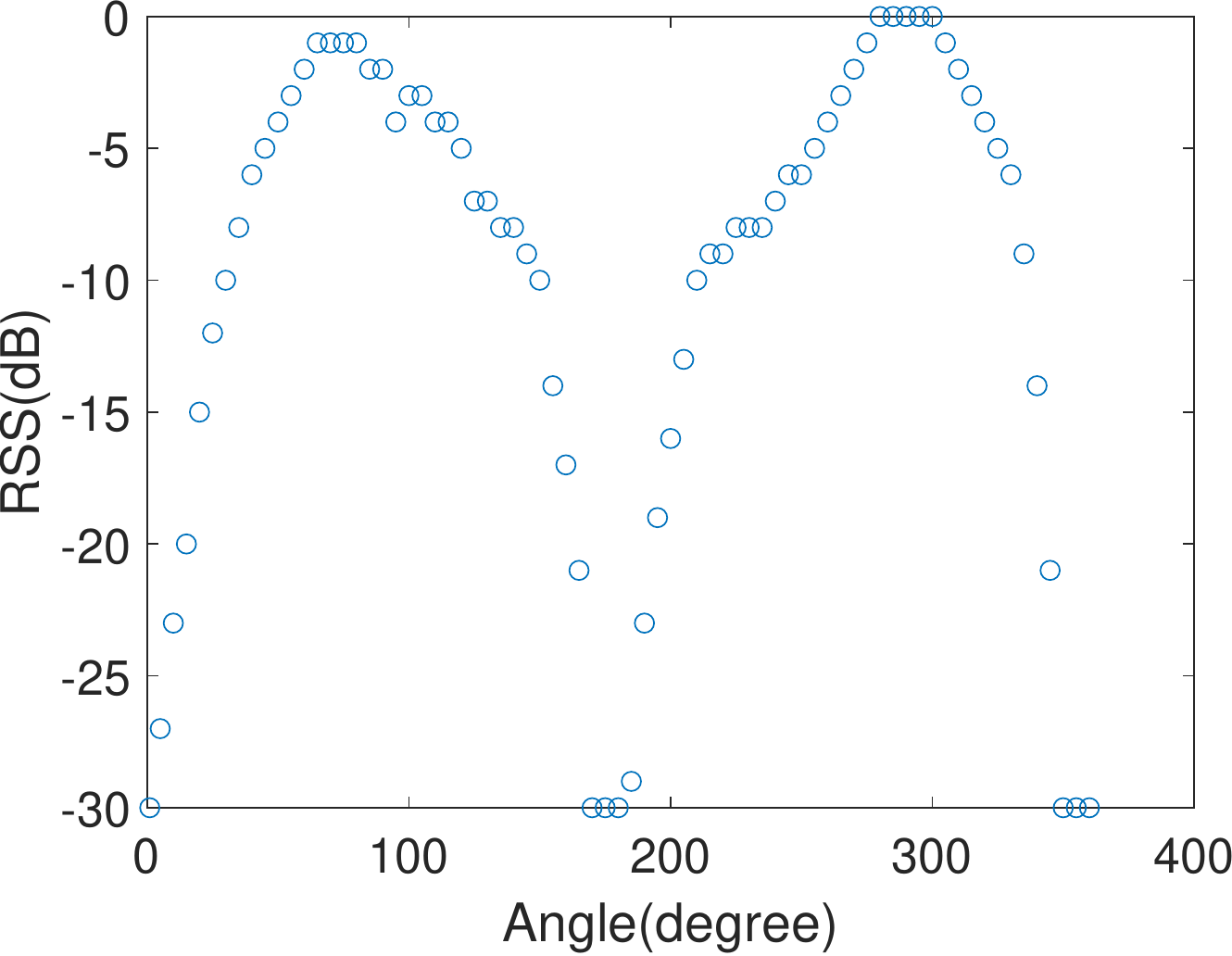}\vspace{-1mm}
    \caption{Vertical radiation pattern for the BroadSpec antenna at 4~GHz. 
    }\label{Fig:AntennaSpec}\vspace{-5mm}
\end{wrapfigure}

The antenna gains $G_{\rm TX}(\alpha)$ and $G_{\rm RX}(\alpha)$ in~\eqref{Eq_1} can be determined based on the vertical or horizontal orientation of the antennas at the transmitter and receiver. In one approach, the antenna gains can be extracted from the data sheets as provided in Fig.~\ref{Fig:RadiationPattern} from~\cite{BroadSpecAntenna}. In Fig.~\ref{Fig:AntennaSpec}, as an example, we plot the vertical antenna gain as a function of the elevation angle for a central frequency of 4~GHz.



     
    


\begin{table}[h!]
\centering
 \caption{Antenna gains with the considered model for different elevation angles and antenna orientations.}\label{Tab1}
 \begin{tabular}{|c||c|c|c|} 
 \hline  & {\bf VV} & {\bf VH} & {\bf HH} \\ 
 \hline  $G_{\rm RX}(\alpha)$ & $\cos(\alpha)$ & $\cos(\alpha)$ & $\sin(\alpha)$ \\ 
 \hline  $G_{\rm TX}(\alpha)$ & $\cos(\alpha)$ & $\sin(\alpha)$ & $\sin(\alpha)$ \\ 
 \hline  $G_{\rm RX}(\alpha)G_{\rm TX}(\alpha)$ & $\cos^2(\alpha)$ & $0.5\sin(2\alpha)$ & $\sin^2(\alpha)$ \\ 
\hline
 \end{tabular}
\end{table}

As an alternative to using the antenna gains specified in the manufacturer's data sheets, in this paper we consider a simplistic model as illustrated in Fig.~\ref{Fig:GeometricModel} to characterize the antenna gain. In particular, we consider that the radiation pattern can be approximated by a circle in the vertical dimension, and it is constant for all horizontal directions. Then, for the three antenna orientation scenarios in Fig.~\ref{Fig:Orientations}, gains in~\eqref{Eq_1} can be calculated easily as summarized in Table~\ref{Tab1}. For example, for the VV configuration in Fig.~\ref{Fig:Orientations}, $G_{\rm RX}(\alpha)=G_{\rm TX}(\alpha)\rightarrow 1$ when $\alpha\rightarrow 0$, which happens when the horizontal distance between the transmitter and the receiver increases, and $G_{\rm RX}(\alpha)=G_{\rm TX}(\alpha)\rightarrow 0$ when $\alpha\rightarrow 90^\circ$ which happens when the drone moves over the ground receiver. Similar observations can be made for the VH and HH antenna orientations. 

For the VV scenario, the horizontal distance that maximizes the RSS at the ground receiver can be derived easily. Letting $l$ denote the horizontal distance between the ground station and the drone, and collecting all fixed coefficients in a variable $K$, we can write the derivative of \eqref{Eq_1} with respect to $l$ as
\begin{align}
\frac{\partial P_{\rm RX}(\alpha)}{\partial l} &= \frac{\partial}{\partial l} \cos^2\bigg(\arctan\frac{h}{l}\bigg)\frac{K}{(l^2+h^2)^{\gamma/2}}\nonumber\\
&= \frac{2h^2 - l^2\gamma}{l(h^2+l^2)^{0.5\gamma-2}}~,\label{Eq:CrDist}
\end{align}
which we can equate to zero to solve for horizontal distance that maximizes the RSS as $l=\sqrt{2h^2/\gamma}$, which for a path loss exponent of $\gamma=2$ gives $l=h$ as the critical distance. A similar analysis can be carried out for the VH scenario 



\subsection{Using Multiple Antennas for Improved Connectivity}\label{Sec:MultiAnt}

\begin{wrapfigure}{r}{2.1in}
\centering \vspace{-4mm}
    \includegraphics[width=2.1in]{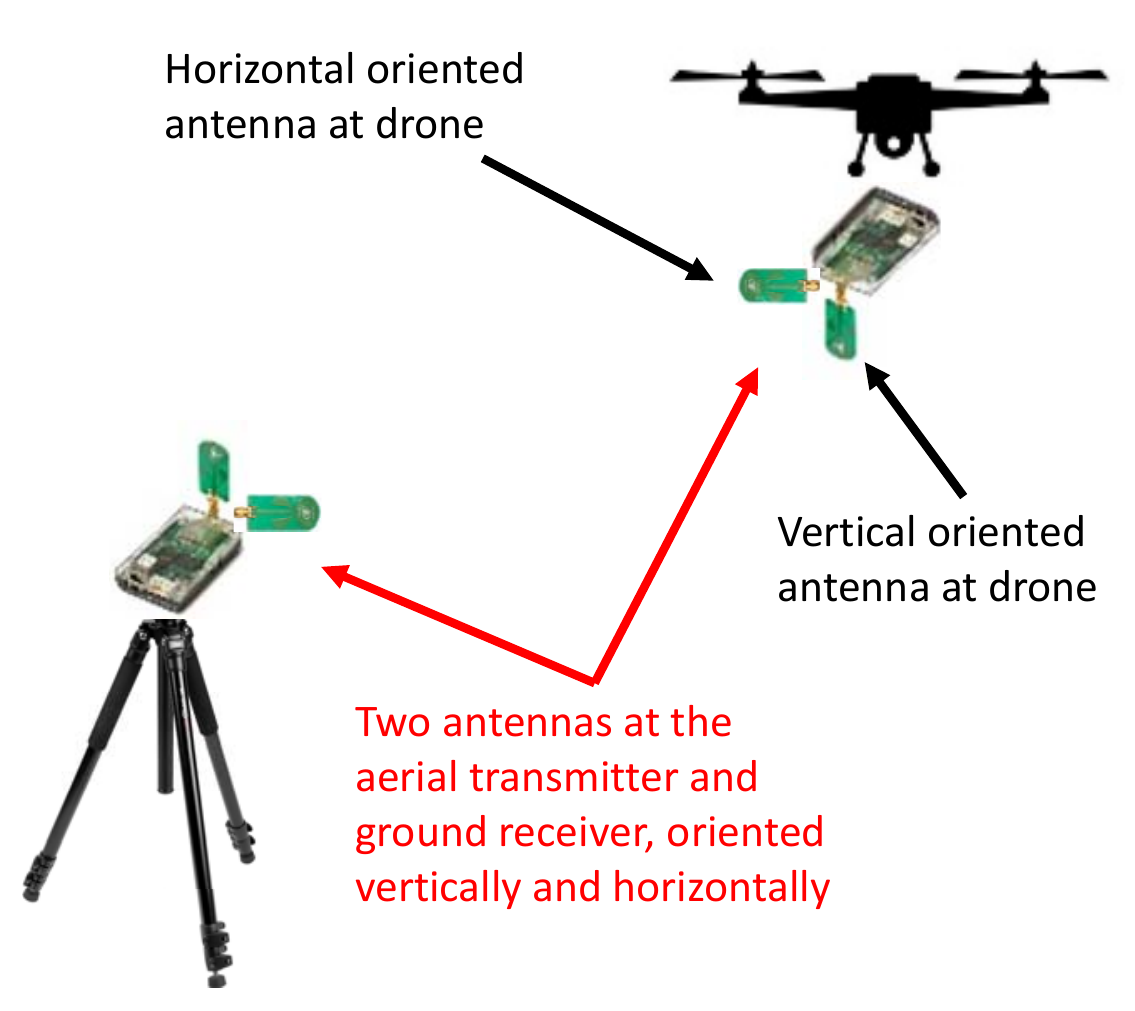}\vspace{-1mm}
    \caption{Two antennas at the aerial transmitter and the ground receiver.}\label{Fig:MIMO}\vspace{-4mm}
\end{wrapfigure}

In this section we present a framework to analyze coverage improvements using multiple different sets of antennas for A2G links. In particular, we consider a simple scenario as in Fig.~\ref{Fig:MIMO} where two UWB  antennas are used at both the aerial transmitter (simply sending the same signal) and the ground receiver,  with one oriented vertically and the other oriented horizontally. Then, considering only the line of sight propagation, the total gain experienced by the horizontally positioned ground receiver antenna is given by: 
\begin{align}
G_{\rm RX-H}(\alpha)=\sin^2(\alpha) +\cos(\alpha)\sin(\alpha)~,
\end{align}
while the gain at the vertical ground receiver antenna is:
\begin{equation}
G_{\rm RX-V}(\alpha)=\cos^2(\alpha) +\cos(\alpha)\sin(\alpha)~.
\end{equation}
In particular, we consider that the total gain at the vertically oriented receive antenna is the sum of the gains due to the VV and VH components and the total gain at the horizontally oriented receive antenna is the sum of the gains due to the HH and HV components. If we consider a simple antenna selection technique at the ground receiver to choose the strongest signal, we can write:  
\begin{equation}
G_{\rm RX-max}(\alpha)=\max\Big(G_{\rm RX-H}(\alpha),G_{\rm RX-V}(\alpha)\Big).
\end{equation}
Obviously, more complicated multiple antenna transceiver techniques can be considered, but we will limit our discussion to simple antenna selection with only two transmit/receive antennas to study achievable gains with a commonly used (doughnut-shaped) antenna radiation pattern in 3D A2G links.

\section{Numerical and Experimental Results}

\begin{figure}[h!]
  \centering
  \begin{subfigure}[b]{0.75\columnwidth}
 \hspace{-0.65cm} \includegraphics[width=3in]{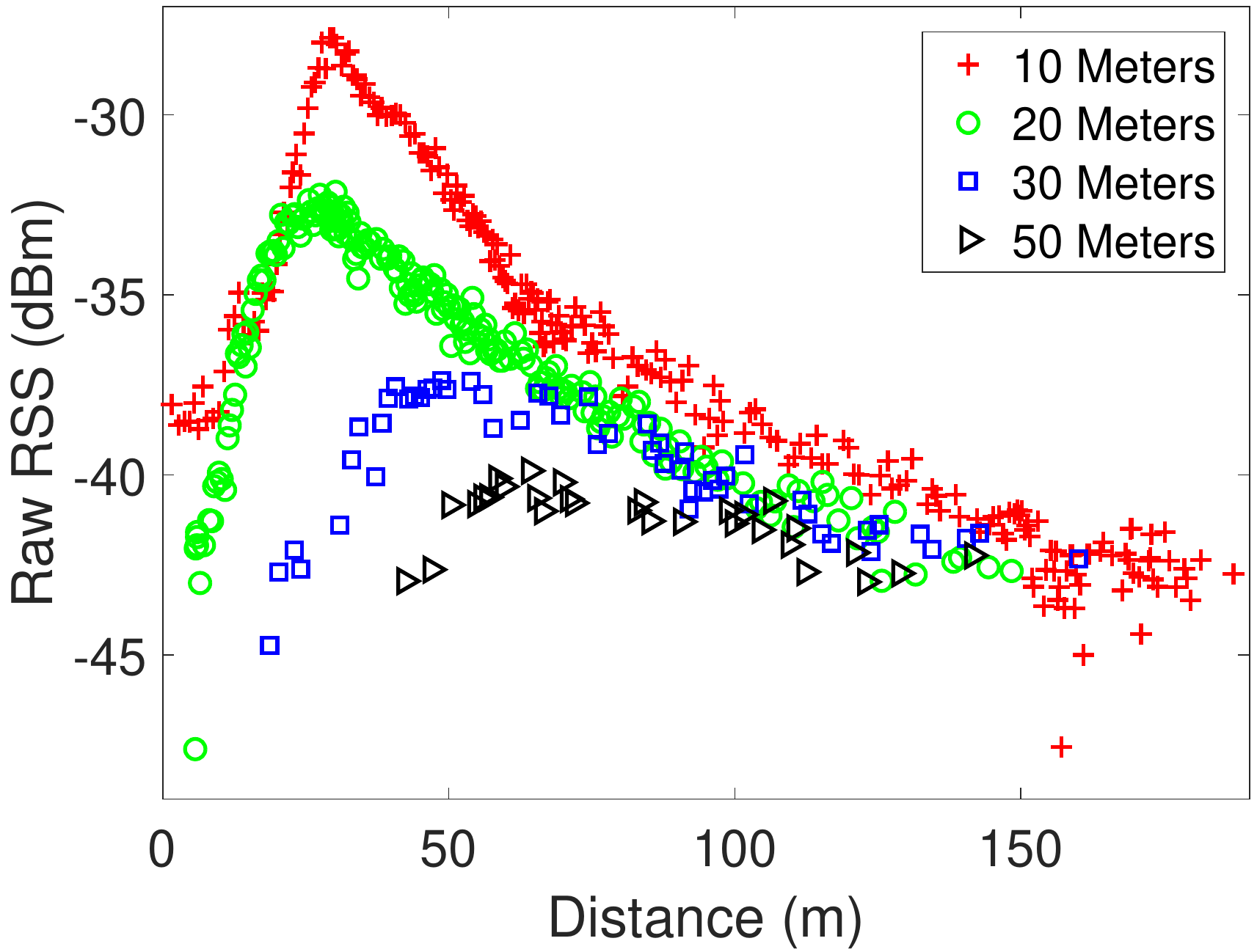}\caption{VV antenna configuration.} 
  \end{subfigure}\\
    \begin{subfigure}[b]{0.75\columnwidth}
 \hspace{-0.65cm}  \includegraphics[width=3in]{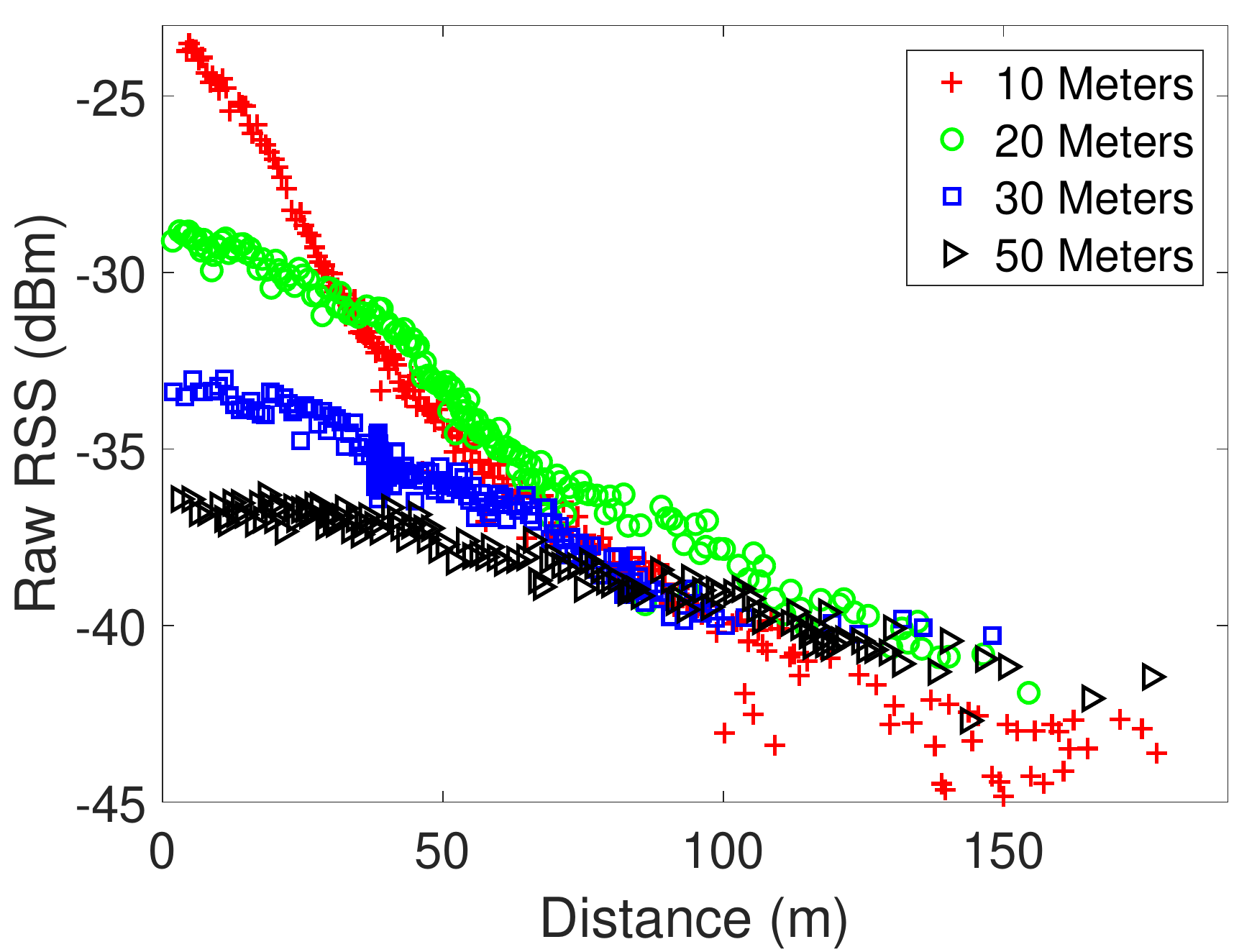}\caption{HH antenna configuration.} 
    \end{subfigure}\\
    \begin{subfigure}[b]{0.75\columnwidth}
 \hspace{-0.65cm}  \includegraphics[width=3in]{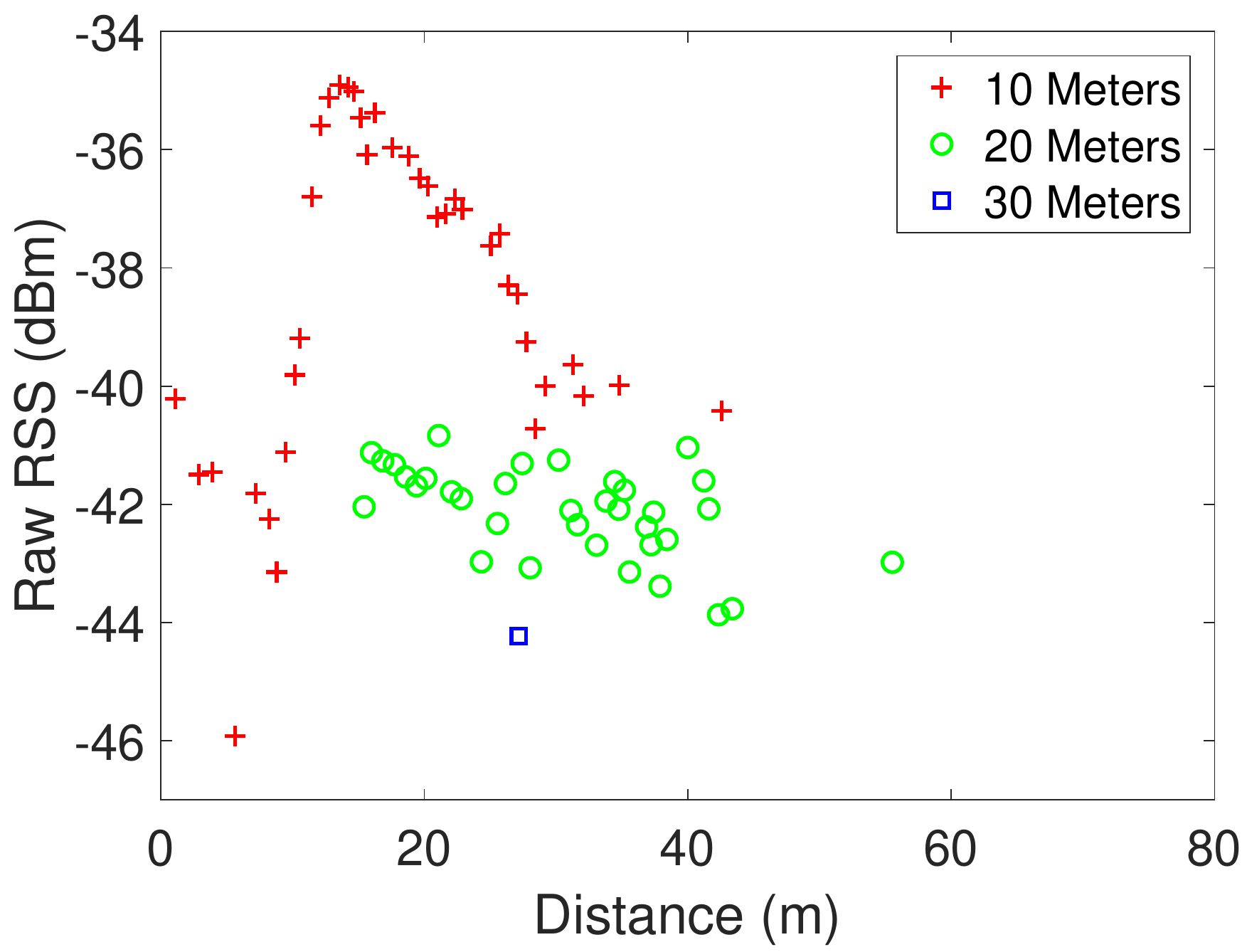}\caption{VH antenna configuration.}  
    \end{subfigure}
    \vspace{-1mm}
  \caption{Raw RSS measurements at four different drone heights for (a) VV, (b) HH, and (c) VH antenna orientations.}\label{Fig:Raw_RSS_measurements}\vspace{-5mm}
\end{figure}

  

In this section we present numerical and experimental results based on the measurement scenarios in Fig.~\ref{Fig:DroneEnvir} and Fig.~\ref{Fig:Orientations}. A P440 UWB radio is attached to the drone with different antenna orientations, and the drone flies in a linear path at a height of 10~m, 20~m, 30~m, and 50~m, up to a horizontal distance of 200~m.  The RangeNet software for P440 radios is used to estimate the distance traveled, and to capture the RSS samples at a ground receiver. 

\begin{figure*}[t]
\centering 
	\begin{multicols}{2}
    \begin{subfigure}{0.5\textwidth}
	\includegraphics[width=3in]{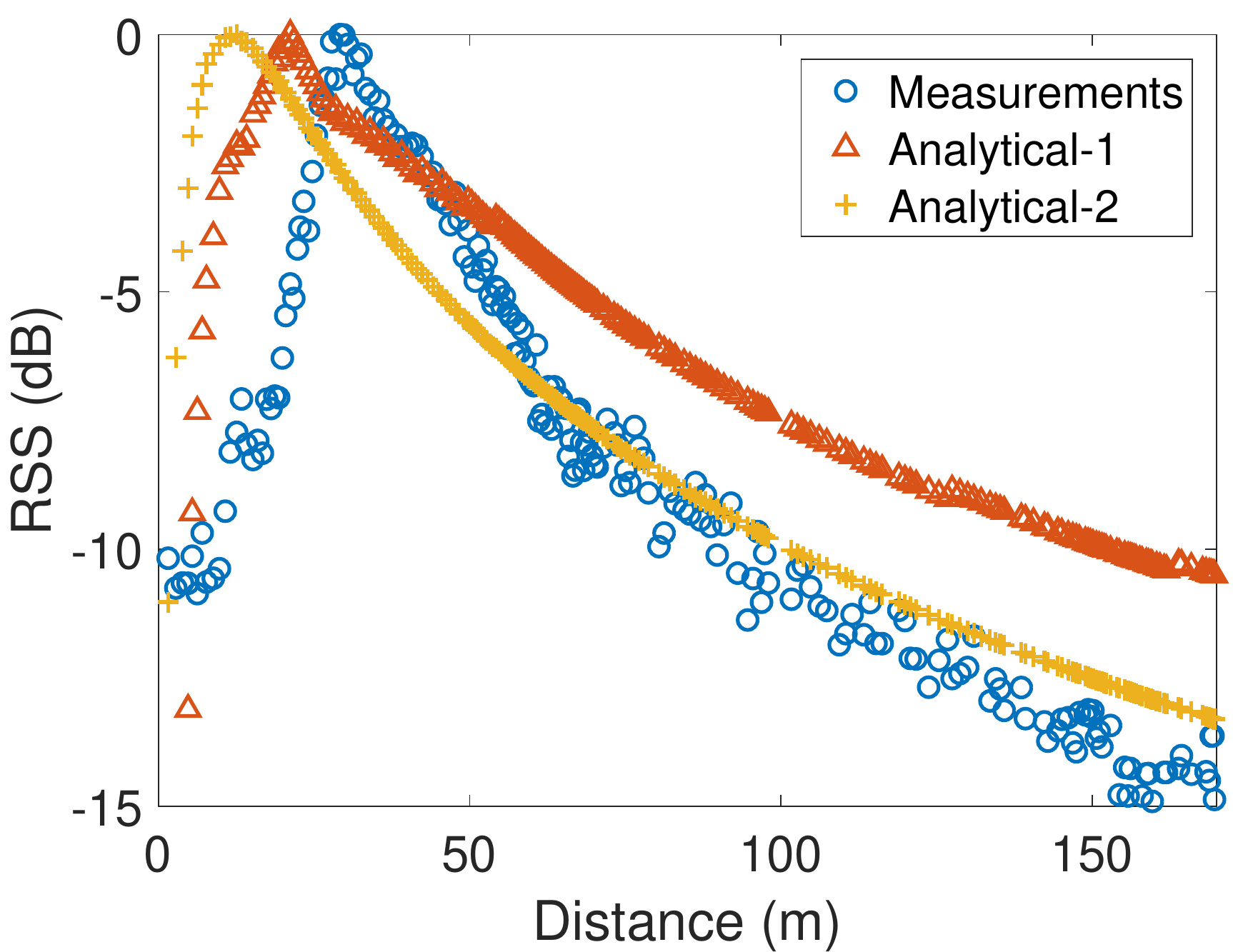} 
    \caption{Drone height: 10 meters.}
	\end{subfigure}
 
	\begin{subfigure}{0.5\textwidth}
    \includegraphics[width=3in]{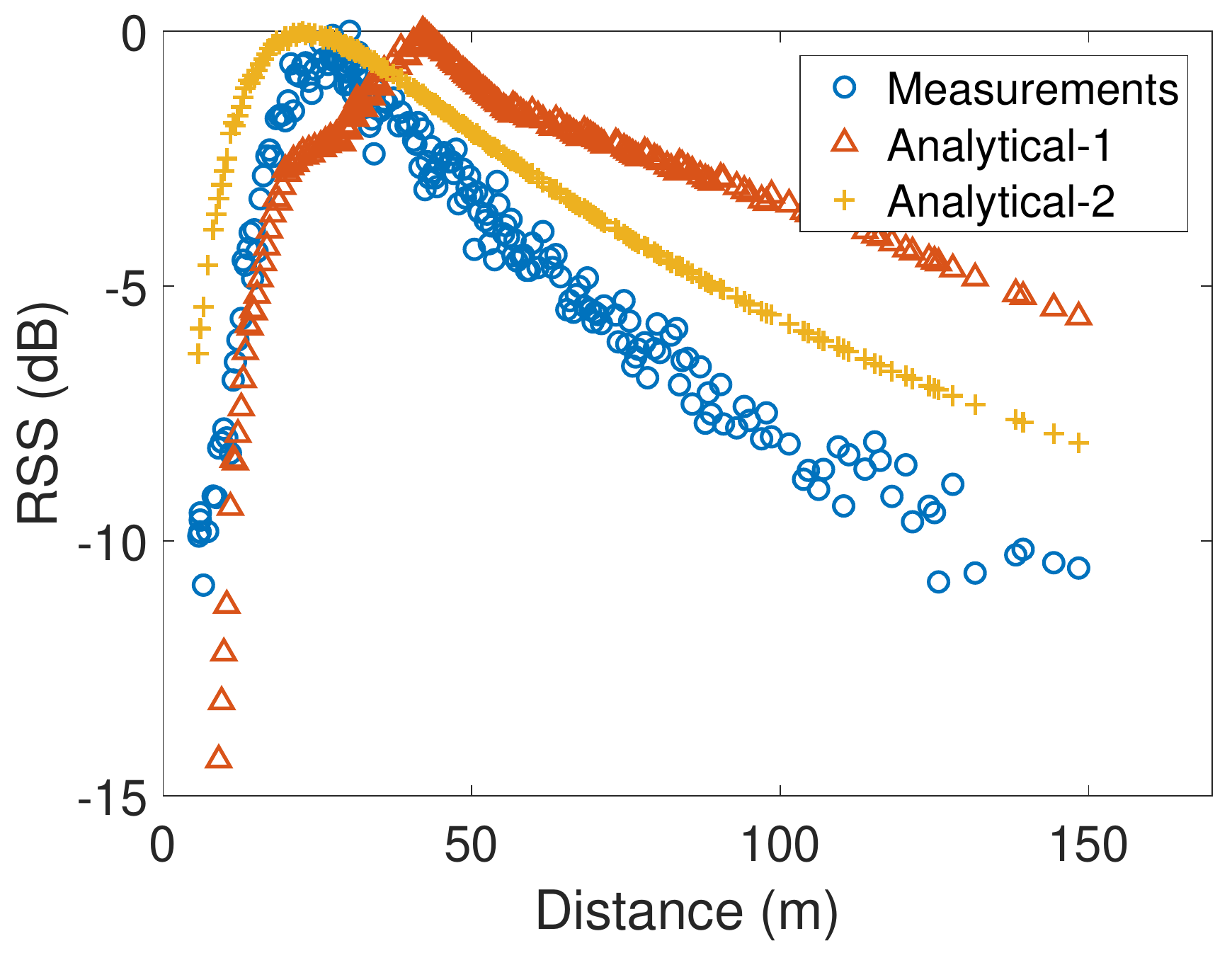} 
       \caption{Drone height: 20 meters.}
    \end{subfigure}
    \end{multicols} 
    
    \begin{multicols}{2}
    \begin{subfigure}{0.5\textwidth}
	\includegraphics[width=3in]{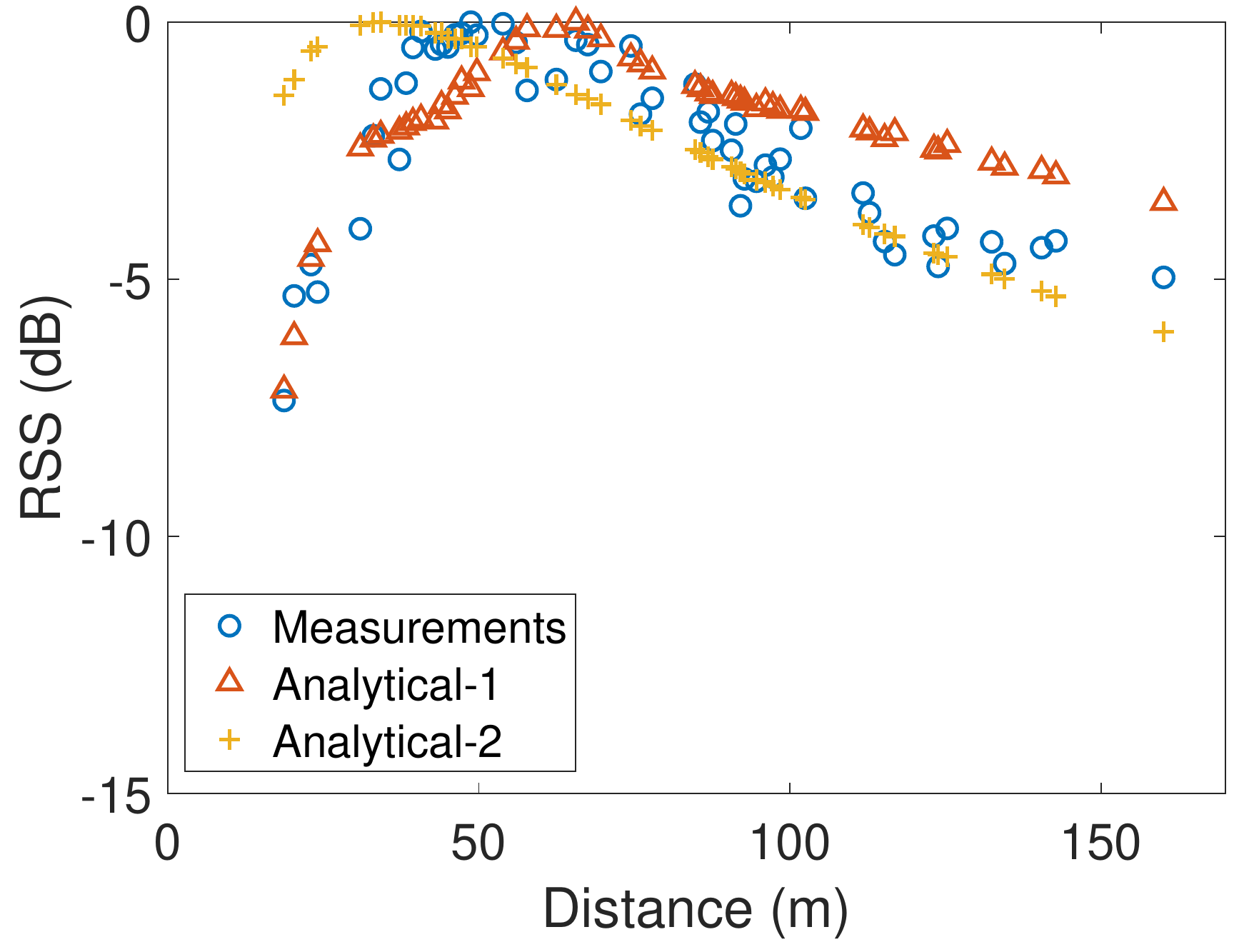} 
       \caption{Drone height: 30 meters.}
	\end{subfigure}

	\begin{subfigure}{0.5\textwidth}
    \includegraphics[width=3in]{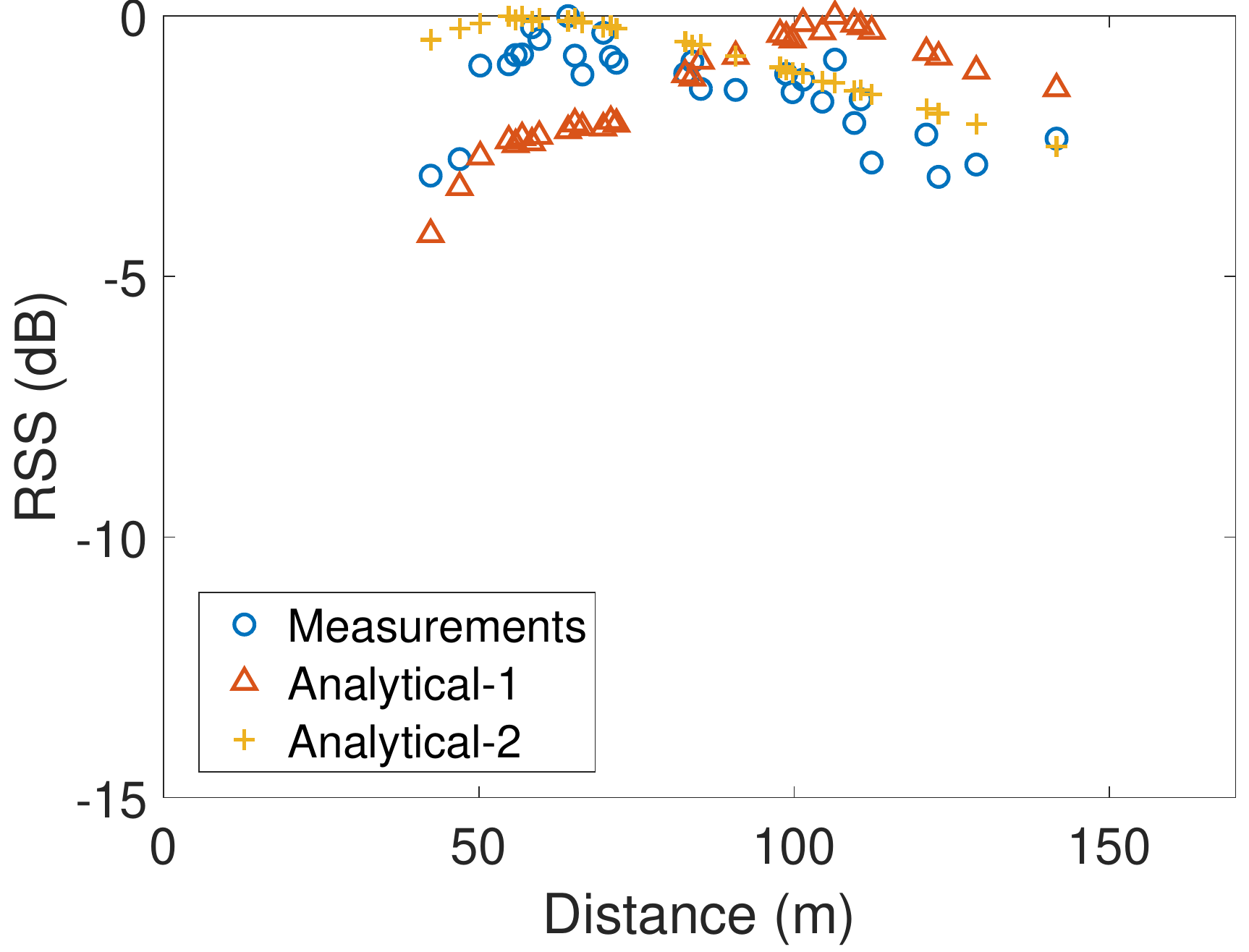} 
       \caption{Drone height: 50 meters.}
    \end{subfigure}
    \end{multicols} 
\vspace{-2mm}
\caption{Normalized RSS results for the VV scenario considering drone heights of 10~m, 20~m, 30~m, and 50~m.}\label{Fig:VV_measurements}\vspace{-4mm}
\end{figure*}

\subsection{Raw Measurements}

First, we provide our raw RSS measurement data (in dBm) in Fig.~\ref{Fig:Raw_RSS_measurements} at different drone heights for VV, HH and VH antenna orientations. As can also be predicted from Table~\ref{Tab1}, for the VV and VH scenario the raw RSS increases with increasing distance till the critical distance, and then it starts to decrease, whereas for the HH scenario the raw RSS decreases monotonically with increasing distance. For the VV scenario, for a given drone height, there exists an optimum drone distance that maximizes the RSS at the ground receiver, and this optimum distance is observed to be increasing with increasing drone altitude. The VH configuration is observed to yield the worst RSS, with no RSS recording observed for 50 meters drone altitude (due to RSS being below the sensitivity threshold).

\subsection{Analytical Results with Single Antenna}

We present two sets of analytical results. For \emph{Analytical-1}, the antenna gains for different elevation angles are extracted from  
Fig.~\ref{Fig:RadiationPattern}, 
while \emph{Analytical-2} calculates the antenna gains based on the simple model presented in Fig.~\ref{Fig:GeometricModel}. RSS values for measurements and analytical calculations are \emph{normalized with the peak RSS value} for a given scenario. 

\begin{figure*}
  \begin{multicols}{2}
  \begin{subfigure}{0.5\textwidth}
  \includegraphics[width=3in]{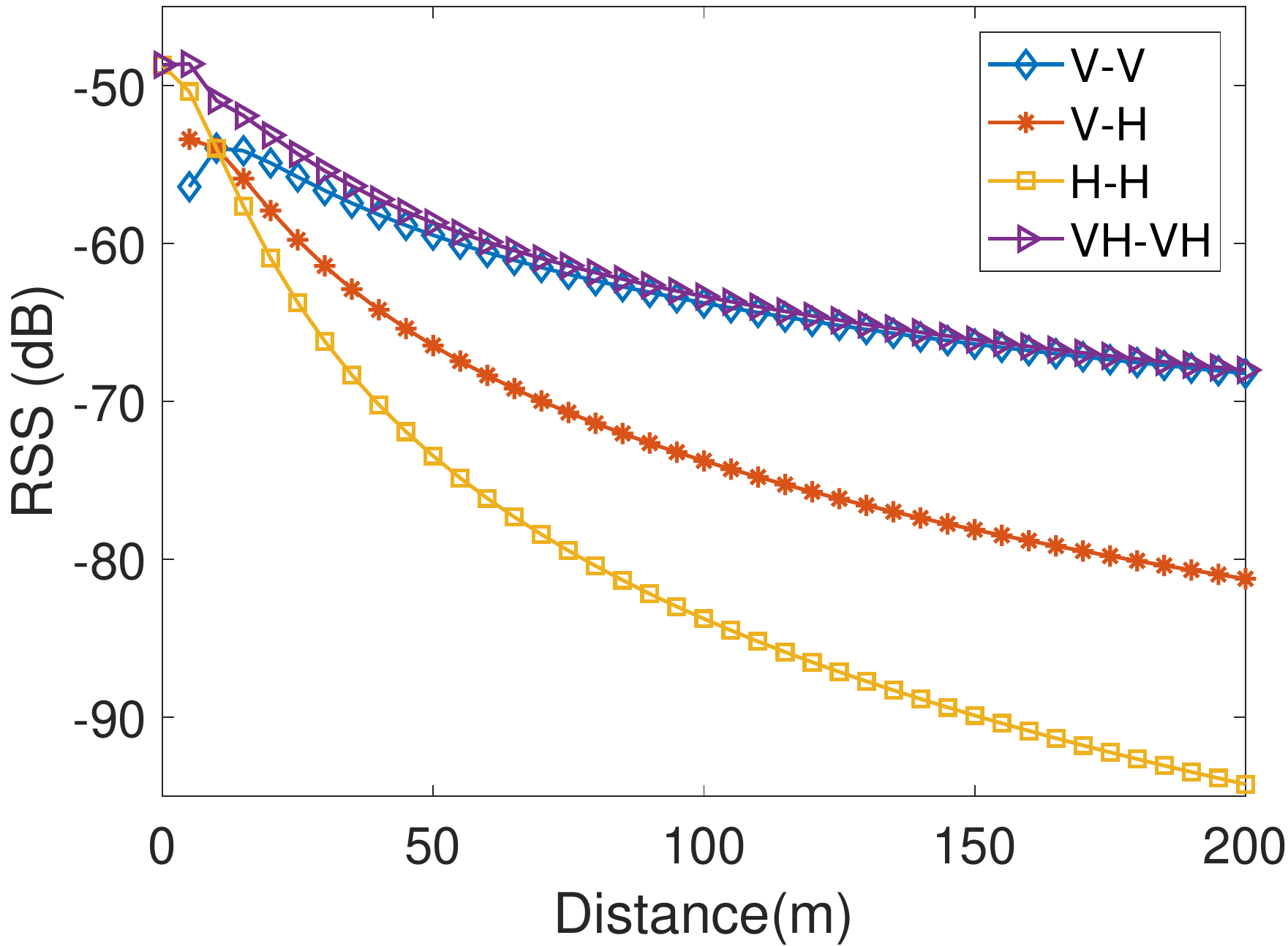} 
   \caption{Drone height: 10 meters.}
  \end{subfigure}
  
    \begin{subfigure}{0.5\textwidth}
\includegraphics[width=3in]{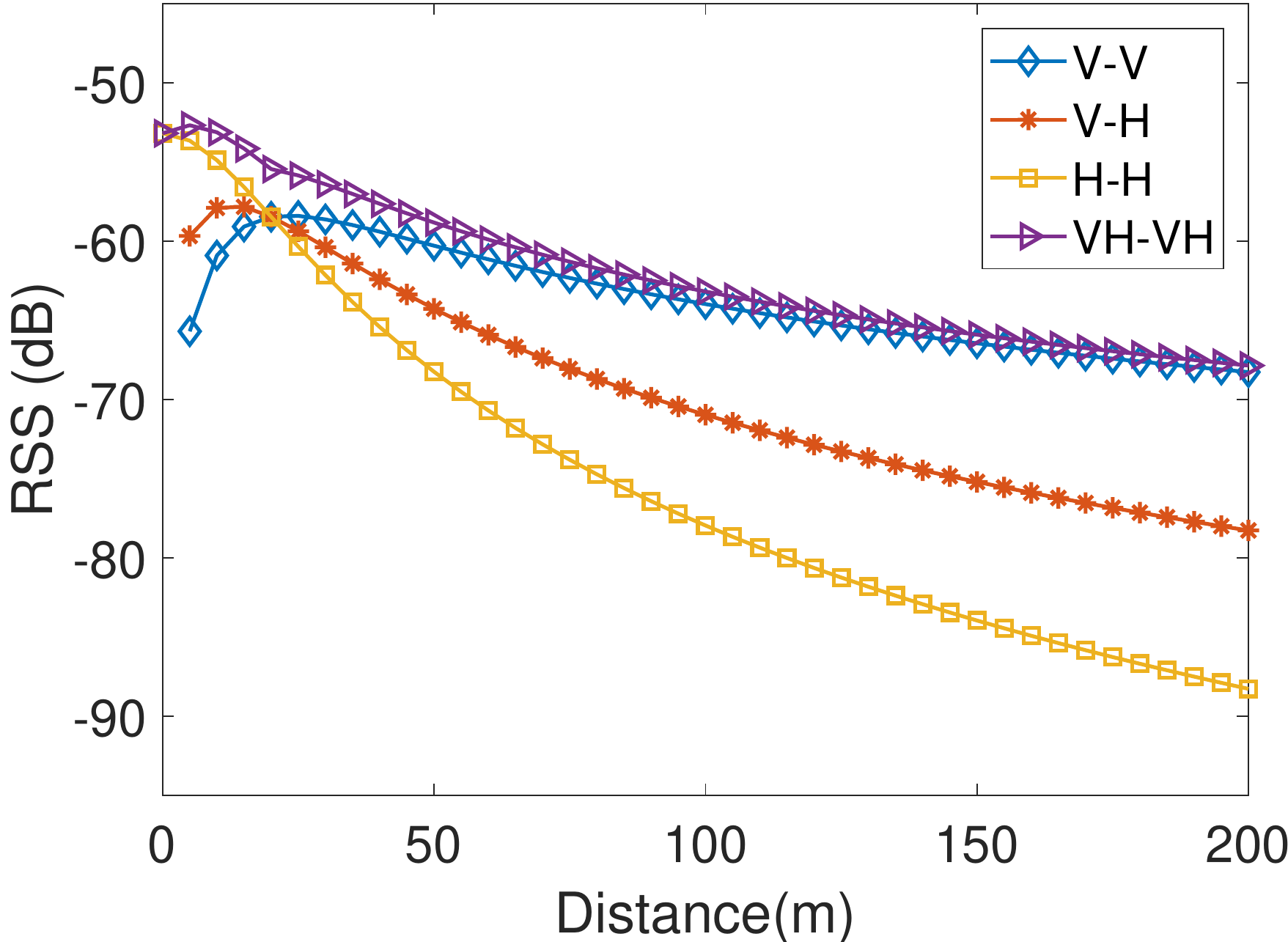} 
   \caption{Drone height: 20 meters.}
  \end{subfigure}
  \end{multicols}
  
  \begin{multicols}{2}
    \begin{subfigure}{0.5\textwidth}
  \includegraphics[width=3in]{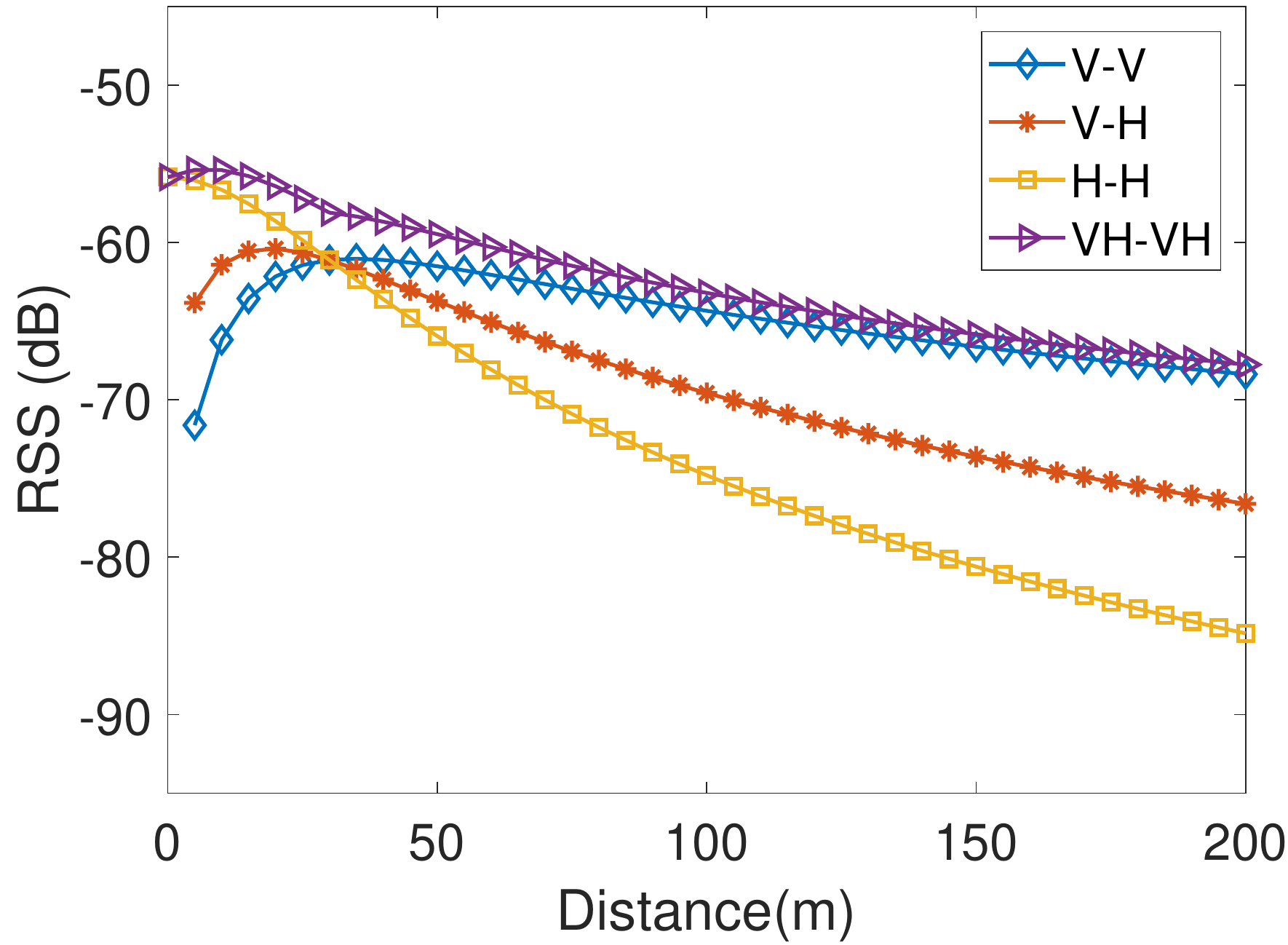} 
  \caption{Drone height: 30 meters.}
  \end{subfigure}
  
  \begin{subfigure}{0.5\textwidth}
  \includegraphics[width=3in]{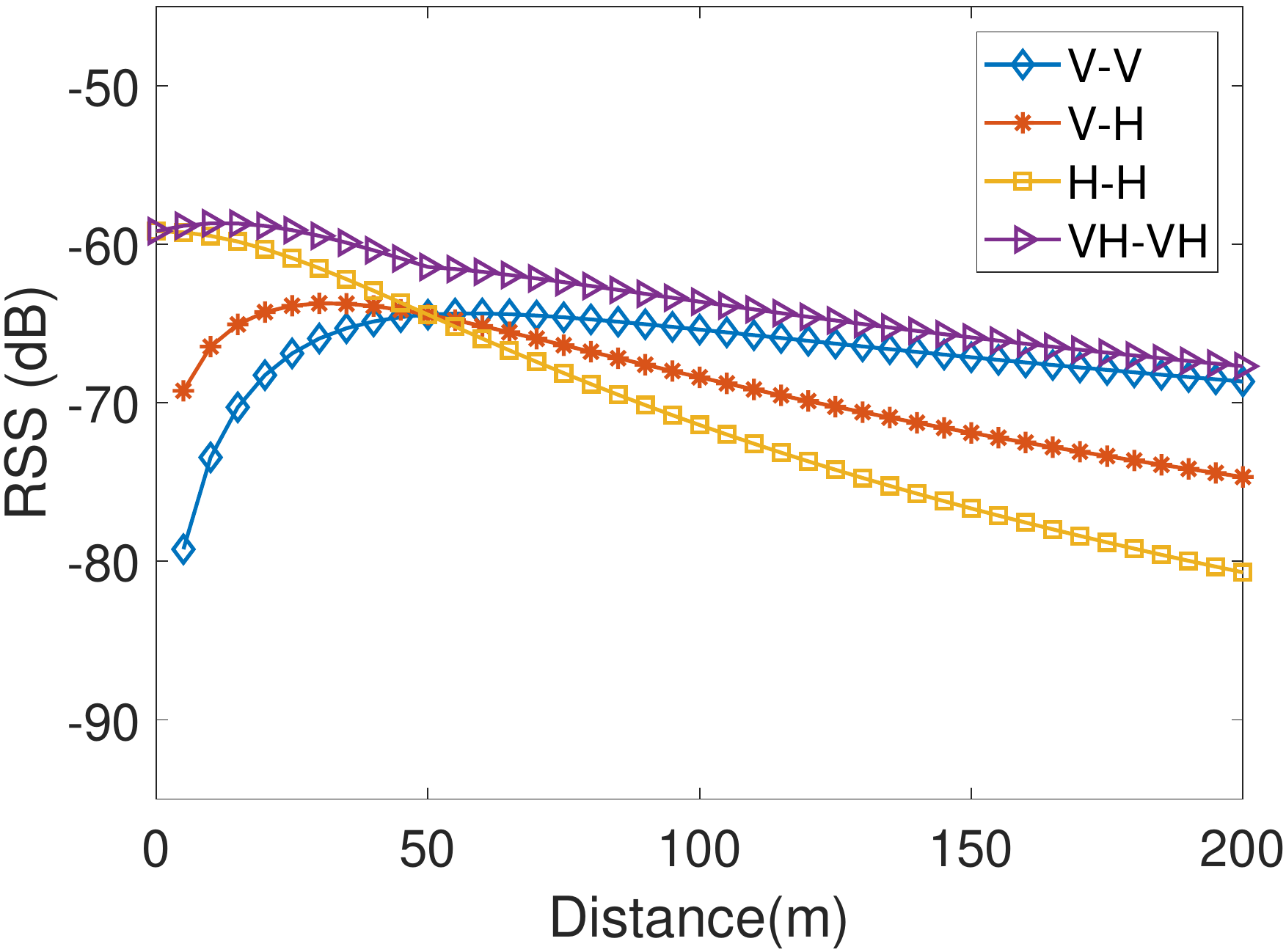} 
  \caption{Drone height: 50 meters.}
  \end{subfigure}
\end{multicols}
 \vspace{-2mm}
  \caption{Analytical RSS versus distance for VV, HH, VH, and VH-VH scenarios at four different drone heights.}\label{Fig:VH-VH Analysis}\vspace{-4mm}
\end{figure*}

Results in Fig.~\ref{Fig:VV_measurements} show the normalized RSS for four different drone heights as a function of the horizontal distance between the drone and the ground station. Measurement and analytical results all show that for the VV scenario, there is an optimum drone to ground station distance where the RSS is maximized. The reason for this is that the antenna gain product $G_{\rm RX}(\alpha)G_{\rm TX}(\alpha)$ for VV scenario increases monotonically as a function of distance, while the path loss also increases with the distance; after a critical distance (which is seen to be aligned with our findings in~\eqref{Eq:CrDist}), path loss dominates the antenna gain, and the RSS starts reducing. 
Finally, larger drone heights result in worse connectivity, as apparent from Fig.~\ref{Fig:VV_measurements}(d) for a height of 50~m where there is no connectivity except for the distance range between 40~meters to 150~meters. 

While the trends for the measurements and analytical results are similar, we observe a mismatch among all the three sets of results, which can be attributed to several possible reasons. First, we used  the antenna pattern centered at 4~GHz for our analytical results, while the overall frequency band for P440 UWB transmissions extend from 3.1~GHz to 5.3~GHz, for which the antenna radiation pattern is not constant. Second, the circular approximation in our analytical model in Fig.~\ref{Fig:GeometricModel} is not aligned perfectly with the exact radiation pattern in Fig.~\ref{Fig:RadiationPattern}. Third, we have not accounted for the horizontal radiation patterns in Fig.~\ref{Fig:RadiationPattern}(b), which can show variations at different frequencies. And finally, measurements and analytical results may deviate from each other due to additional attenuation that may be caused by drone components, occasional random motion of the drone (e.g. due to wind gusts), effect of scatterers/multipath, and other possible impairments. 

\subsection{Analytical Results with Multiple Antennas}

Finally, we use the analytical model in Section~\ref{Sec:MultiAnt} to study how the use of multiple antennas can improve coverage for A2G links with two transmit and two receive antennas as in Fig.~\ref{Fig:MIMO}. Results in Fig.~\ref{Fig:VH-VH Analysis} show how the RSS (normalized with respect to transmit power $P_{\rm TX}$ in ~\eqref{Eq_1}) varies with increasing distance between the transmitter on the drone and the receiver unit on ground, for the three single-antenna based orientation configurations (VV, VH, and HH) and the multiple-antenna based orientation configuration (VH-VH).
With the VH-VH configuration, we are able to maintain a steady and reasonably good link quality at all distances between 0 to 200 meters, and thus overcome the shortcomings of the single-antenna   configurations with different antenna orientations. 
As expected, the VV configuration performs best at large A2G horizontal link distances, the HH configuration performs best at very short A2G link distances, and VH configuration provides a compromise bet ween VH and VV configurations. On the other hand, VH-VH configuration with receiver antenna selection  performs the best among all link distances, due to utilizing the best antenna corresponding to each drone distance. 


\section{Conclusion}
In this paper we presented our findings from a UWB air-to-ground measurement campaign conducted at NCSU for three different UWB antenna orientation configurations, and for various different UAV heights and transmitter-receiver distances. Our simple analytical model can capture the non-monotonic trend in the RSS as a function of transmitter-receiver distance. We also provide insights on improving coverage with multiple different sets of antennas with different orientations. Our future work includes theoretical study of drone outage probability considering 3D antenna radiation.



\bibliographystyle{IEEEtran}


\end{document}